\documentclass[10pt,a4paper]{article}

\usepackage{amsmath}
\usepackage{amssymb}
\usepackage{graphicx}
\usepackage{url}
\usepackage{comment}
\usepackage[numbers,sort&compress]{natbib}

\usepackage{setspace}

\usepackage[left=1in,right=1in,top=1in,bottom=1in]{geometry}

\title{\bf A semantic-based deep learning approach for mathematical expression retrieval}
\date{}

\author{\bf {Pavan Kumar Perepu}\\Indian Institute of Information Technology, Sri City, India\\Email: {pavan.ppkumar@gmail.com}}

\begin{document}

\maketitle

\begin{abstract}
Mathematical expressions (MEs) have complex two-dimensional structures in which symbols can be present at any nested depth like superscripts, subscripts, above, below  etc. As MEs are represented using LaTeX format, several text retrieval methods based on string matching, vector space models etc., have also been applied for ME retrieval problem in the literature. As these methods are based on syntactic similarity, recently deep learning approaches based on embedding have been used for semantic similarity.  In our present work, we have focused on the retrieval of mathematical expressions using deep learning approaches.  In our approach, semantic features are extracted from the MEs using a deep recurrent neural network (DRNN) and these features have been used for matching and retrieval. We have trained the network for a classification task which determines the complexity of an ME. ME complexity has been quantified in terms of its nested depth. Based on the nested depth, we have considered three complexity classes of MEs: Simple, Medium and Complex. After training the network, outputs just before the the final fully connected  layer are extracted for all the MEs. These outputs form the semantic features of MEs and are stored in a database. For a given ME query, its semantic features are computed using the trained DRNN and matched against the semantic feature database. Matching is performed based on the standard euclidean distance and top $k$ nearest matches are retrieved, where $k$ is a user-defined parameter. Our approach has been illustrated on a database of 829 MEs.
\end{abstract}

{\bf keywords.} Mathematical expression retrieval, Deep learning, Embedding, Recurrent neural network, Semantic similarity.

\section{Introduction}
\label{intro}
Mathematical expressions (MEs) form an essential part of the scientific and engineering documents. Scientific languages like LaTeX, MathML etc., can be used to encode and represent MEs. Given a mathematical expression in an encoding language, ME retrieval algorithm matches it against a database and extracts similar expressions. As MEs have complex two-dimensional structures with symbols at different nested depths, matching process should also consider the structure in addition to the symbol identities. For example, in $a^{2^n}_m$, superscripts and subscripts, $2$, $n$ and $m$ are at nested depths, 1, 2 and 1 respectively, with respect to the baseline symbol, $a$. Similarly, symbols can be nested in the above and below regions of fraction expressions, enclosed in squareroot expressions etc. 

We have earlier applied longest common subsequence (LCS) algorithm \cite{ppkretrieval} for matching and retrieval of similar LaTeX expressions. As LaTeX expressions cannot be directly matched, we have converted them into an integer encoding format in which LaTeX keywords are mapped to their corresponding integers. We have created a vocabulary of LaTeX keywords and each keyword is mapped to a unique integer. Matching has been performed on the integer encoded strings. As LCS based algorithm has been used, it has quadratic time complexity. To reduce time complexity, we have applied parallel LCS algorithm using OpenMP framework \cite{ppkppl}.

In recent years, deep learning approaches have been used for text retrieval. In these approaches, text words are embedded into vector space to capture semantic similarity \cite{mikolov2013_1, mikolov2013_2, glove2014, fasttext2017}. Similar approaches can also be applied to compute semantic similarity of MEs. In our present work, we have focused on deep learning approaches for retrieval of MEs based on semantic similarity. We have also compared this deep learning based semantic approach with our earlier LCS based syntactic approach \cite{ppkretrieval}. To the best of our knowledge, this is the first time, semantic and syntactic approaches have been analyzed and compared for ME retrieval problem. Our contributions in this paper are listed below.
\begin{enumerate}
\item A simple and intuitive deep learning architecture for ME retrieval. As discussed later, this architecture considers variable-length input MEs without the need of padding them to a fixed size.
\item Semantic features for MEs are computed using a deep recurrent neural network (DRNN). That means, DRNN has been used for semantic feature extraction from MEs.
\item Matching and retrieval are performed based on the semantic features. 
\item As semantic feature vectors obtained from the DRNN are of fixed size, a standard euclidean distance has been used for their matching. Further, time complexity for the euclidean distance computation is linear.
\end{enumerate}
The rest of the paper is organized as follows: Section \ref{relatedwork} focuses on the existing approaches for ME retrieval. In Section \ref{proposed}, we have presented our deep learning based ME retrieval algorithm. Section \ref{results} discusses the experimental results and the paper is concluded in Section \ref{conc}.

\section{Related work}
\label{relatedwork}

In \cite{recogretrieval, retrieval2016}, authors have presented a detailed survey on ME recognition and retrieval methods. In \cite{mansouri2019tangent}, various math retrieval approaches based on text and tree models have been discussed. In text based approaches \cite{mathuser, mathfind, mathgo}, linear encoding forms like LaTeX have been used and retrieval is based on \emph{bag of words} and vector space models. Tree models \cite{searchmath, graftree, improvemath, mathmlsearch} have focused on matching mathematical tree structures like MathML, which are computationally intensive. In \cite{mathgo,mathcluster}, database expressions have been clustered so that a query is matched with the cluster centroids to reduce time complexity. In \cite{mansouri2019characterizing}, authors have focused on query refinement based on the history and nature of the math queries.

Few approaches \cite{zanimage, mathspot}  are based on content-based image retrieval, where queries are taken in the form of images. In these approaches, an ME image is taken as input, segmented, features (like density, contours etc.) are extracted and matched. A Tangent search engine \cite{zanibbi2015math, zanibbi2016multi, stalnaker2015math} has been proposed based on layout of symbol pairs, has been presented. This search engine creates an inverted index for each pair of symbols and MEs are retrieved based on the matched symbol pairs in the query and database expressions. This interface has also been applied on the image based handwritten queries. In \cite{mathretpdf2020}, PDF mathematical documents have been converted to images and features are computed on the images. Based on these features, a support vector machine classifier has been trained to classify and retrieve the MEs. 

Recently, word embedding approaches \cite{mikolov2013_1, mikolov2013_2, glove2014, fasttext2017} proposed for text retrieval based on semantic similarity have also been applied for ME retrieval. In \cite{mathembed2016}, authors have focused on the embedding of mathematical concepts and types to retrieve mathematical information from scientific documents. In \cite{Krstovski2018EquationE}, unsupervised methods for semantic representation of MEs based on the surrounding words have been presented. In \cite{Zhong_Zanibbi_2019}, a formula embedding method based on symbol layout and operator trees has been proposed. The authors have used a DeepWalk algorithm to generate embeddings for the trees. Their approach has been tested on NTCIR-12 datasets \cite{ntcir12mathir}. 

 In \cite{mathwordsemantic2020}, text word embedding approaches have been used for various retrieval tasks like math query expansion and similarity, extraction of semantic knowledge etc.  In \cite{semanticsearch2020}, authors have proposed to learn math embedding using graph convolutional neural networks for semantic search of MEs. In \cite{deepmetric2021}, multi-modal image and graph based deep learning models have been used to learn math embeddings. In \cite{zannibi2021}, SVM-rank model has been used for ME retrieval. To train this model, similarity scores from different formula retrieval methods have been used as features. 

 In \cite{bertembedret2021}, formula embedding has been obtained using BERT models and based on cosine similarity between embeddings, top $k$ MEs have been retrieved.  In \cite{treeembed2021}, authors have focused on the generation of embeddings from expression operator trees and vice-versa, which have been used in their formula retrieval framework. Few other approaches \cite{mathyolo2022,  mathmultimodal2023, mathretgraph2023,  formulaembed2023,  mathuse2024} have used complex deep learning architectures like YOLO, graph convolutional networks with fuzzy based similarity computation and transformer based models for math embedding and retrieval. 
 
 \section{Proposed approach}
\label{proposed}
In our present work, we have proposed a simple and intuitive deep learning architecture for feature extraction and retrieval of MEs. We have earlier proposed an LCS based algorithm for ME retrieval \cite{ppkretrieval}. In this approach \cite{ppksadh, ppkretrieval}, LaTeX expressions are preprocessed and converted into an integer encoding format. A vocabulary of LaTeX keywords has been generated and each LaTeX keyword is mapped to a unique integer. LaTeX expressions are scanned and converted to integer format using this mapping table that converts LaTeX keywords to integer values/symbols. Special integer symbols are used to enclose superscript and subscript expressions between special start and end integer symbols recursively. Similarly, special row and column delimiter integer symbols are used for multi-line MEs like matrices. As variable names are not important in the retrieval process, all the variables are mapped to a unique integer. For example, $a^2 + b^2$ and $x^2 + y^2$, should be the same for the retrieval algorithm. Consider an example LaTeX expression: \textbf{\textbackslash sum\_\{i=a\}\^{ }b f(i)} ($\sum_{i=a}^b f(i)$). Its integer encoding form \cite{ppkretrieval} is given by: 102, 1000, 1004, 201, 1004, 1001, 1002, 1004, 1003, 1004, 156, 1004, 157. In the vocabulary, the integers, 102, 201, 156 and 157 correspond to the LaTeX keywords for $\Sigma$, $=$, $($ and $)$ respectively. Start and end integer symbols for superscript (subscript) expressions are denoted by 1000 and 1001 (1002 and 1003) respectively. All the variable names ($i$, $a$, $b$ and $f$) are mapped to a unique integer, 1004. 

All the database MEs in LaTeX format are converted and stored in the integer encoding format. A LaTeX query expression is also converted to the integer encoding form and matched against the integer encoded database using the LCS algorithm \cite{ppkretrieval}. In this approach, matching is based on syntactic similarity of MEs. As LCS based algorithm has been used, time complexity is $O(n^2)$, where $n$ is the number of integer encoded symbols. 

\begin{figure}
\centering
\includegraphics[scale=0.6]{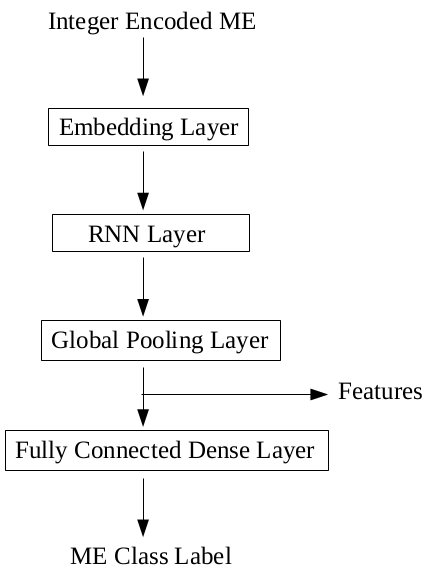}
\caption{Proposed deep learning architecture for ME retrieval.}
\label{drnn}
\end{figure}

Our deep learning approach for ME retrieval is based on semantic similarity of MEs. In this approach, features of MEs are extracted by training a deep learning model on an auxiliary classification task. The classification task can be chosen  based on the user requirements. For example, a user may be interested to retrieve MEs based on their complexity (like Simple, Medium and Complex classes), field (like Algebra, Trigonometry, Geometry and Calculus etc., classes). This deep learning model takes an ME in an integer encoded format as an input, and gives an output classification label (as mentioned above). After training, features are extracted and stored for all the database MEs. As discussed later, features are obtained from the output of the model excluding the final classification layer. A query is also fed to the model and its features are computed. Features of the query are matched against the features of database MEs using the standard euclidean distance and top $k$ MEs are retrieved. Our architecture is shown in Fig. \ref{drnn}.

\subsection{Architecture}
As shown in Fig. \ref{drnn}, proposed deep learning model for ME retrieval has several layers. As discussed earlier, ME in the integer encoded format is given as input to the model. As MEs vary in lengths (number of symbols), model should consider the variable-length inputs. However, fully connected dense layers in any deep learning model \cite{goodfellowdlbook2016}, require fixed-length inputs. To overcome this issue, we have added global pooling layer just before the dense layer. Global pooling layer \cite{goodfellowdlbook2016} transforms its variable-length input to a fixed-length output. Most of the text retrieval approaches \cite{mikolov2013_1, mikolov2013_2} perform padding operation on the model input. For example, in text classification, input sentences may have different number of words. After the integer encoding, a dummy padding symbol (like zero) can be added to all the sentences so that all of them have equal lengths. Number of padding symbols to be added depends on the maximum length of the sentences in the database. If the length of a sentence is $l$ and the maximum length is $m$, then $(m-l)$ padding symbols are added to the sentence. However, this padding operation adds unwanted information or noise to the inputs. As such, we have not used padding and used global pooling operation to handle variable-length inputs. Each layer in our model is discussed below. Here, output of each layer forms the input of the next layer (except the first input layer).

\begin{enumerate}
\item Input to the model is an integer-encoded ME. Let its length be $l$ ($l$ may be different for different MEs). 
\item Embedding operation \cite{mikolov2013_1} is applied on the input ME so that each integer symbol in the input is mapped to a $e$-dimensional vector. If the length of the input ME is $l$, output of this embedding layer is of size, ($l \times e$). Embedding layer serves as look-up table that maps each symbol in the Vocabulary to a $e$-dimensional vector. 
\item All the $l$ embedding vectors (of size, $e$) are fed to a recurrent neural network (RNN) with one layer. If RNN layer has $t$ neurons, $t$ outputs are obtained for each embedding vector. As there are $l$ embedding vectors, output of the RNN layer is of size, ($l \times t$). 
\item As $l$ is a variable, we have added a global pooling layer to obtain fixed size output before applying to the final fully connected dense layer. Input to the global pooling layer is of size, $l \times t$ (output from the previous RNN layer). This layer applies a global pooling (maximum, average or minimum) operation \cite{goodfellowdlbook2016} on each column so that each column is reduced to a single value and so the output is of size, ($1 \times t$). Therefore, this layer generates a fixed $t$-dimensional vector as output, irrespective of the length of the input ME. As shown in Fig. \ref{drnn}, this $t$-dimensional vector forms the features of the given input ME. 
\item The fixed length output from the global pooling layer is fed as input to the final fully connected dense layer. The number of neurons in this layer depends on the number of classes in the auxiliary classification task chosen to train this model. As mentioned earlier, classification can be based on complexity, field etc. 
\end{enumerate}

Here, embedding vector size ($e$), RNN variant (like Vanilla RNN, GRU, LSTM \cite{goodfellowdlbook2016}etc.) and number of neurons in RNN layer ($t$) are the design parameters that can be tuned based on the model performance.
\subsection{Training}
For our earlier works on the recognition and retrieval of MEs \cite{ppkpaa2018, ppkretrieval}, we have created our PACME database of 829 ME images from different fields like Algebra, Trigonometry, Calculus etc., along with corresponding LaTeX ground truth \cite{pacme}. This database consists of different types of MEs, ranging from simple to complex expressions. 

Based on the complexity, we have partitioned this LaTeX database into three classes with the following labels: Simple, Medium and Complex. As discussed in Section \ref{intro}, complexity is quantified in terms of nested depth. Simple expressions do not have nested symbols, Medium expressions have nested symbols with depth one while complex ones have nested depth more than one. Our deep learning model has been trained for this classification task based on complexity. For the sake of illustration, we have chosen this task. However, our proposed model or framework can be used to train for any other task (like classification based on field) based on the user requirements. 

We have divided the database of 829 LaTeX expressions into training and test sets, in the ratio, 70:30. Further, 20\% of the training samples have been used for validation during the training process \cite{goodfellowdlbook2016}. We have implemented the model using Python Keras library \cite{keras2015}. In our model, we have used LSTM variant for the RNN layer. As mentioned earlier, we have trained the model with different values of parameters, $e$ and $t$, and picked up the ones that have shown good performance on both the training and testing samples. In Table \ref{paramdnn}, we have shown accuracies on the training and testing samples for different values of $e$ and $t$. 

\begin{table}
\centering
\caption{Percentage accuracies on training and testing samples ($Acc_{train}$ and $Acc_{test}$) for different values of parameters, $e$ and $t$, of the model.}
\label{paramdnn}
\vspace{0.1cm}
\begin{tabular}{|c|c|c|}
\hline
($e$, $t$) & \% $Acc_{train}$ & \% $Acc_{test}$\\\hline\hline
(16, 32) & 90.95 & 87.95\\ \hline
\textbf{(16, 64)} & {\bf 100} & {\bf 95.58} \\ \hline
(32, 32) & 100 & 91.16\\ \hline
(32, 64) & 99.78 & 92.77\\ \hline
\end{tabular}
\end{table}

Though we have experimented with different values of $e$ and $t$, we have selectively shown a few values that have given a training accuracy more than 90\%, in Table \ref{paramdnn}. From Table \ref{paramdnn}, it can be observed that the model has given the best performance on both the training and testing samples (100\% and 95.58\%, respectively) at $e=16$ and $t=64$. If $t$ is less than 64, performance is low as shown in the first row of the table.  If both $e$ and $t$ values are increased to (32, 32), accuracy on the testing samples is reduced due to overfitting (third row of the table). If $t$ is increased to 64, there is an improvement in the testing accuracy (fourth row). If $t$ is increased further, there may be an improvement in the performance. However, retrieval time is also increased as ultimately $t$-dimensional features are matched during retrieval. As such, we seek a smaller $t$ value that is optimal in terms of performance as well as the retrieval time (high performance and low retrieval time). In the global pooling layer, global pooling based on the minimum operation (among the three operations, minimum, average and maximum) has shown the better performance. 

The final fully connected dense layer has a number of neurons which is equal to the number of classes so that neuron, $i$, corresponds to a class, $i$. Neuron, $i$, gives the probability of class, $i$, using a standard {\bf Softmax} activation function \cite{goodfellowdlbook2016}. If there are $c$ neurons which correspond to $c$ classes, then the probability of each class, $i$, denoted by $P(i)$, is given by Eq. (\ref{softmax}).

\begin{equation}
P(i) = \dfrac{e^{out_i}}{\sum_{k=1}^c e^{out_k}}
\label{softmax}
\end{equation}

Here, $out_k$ is the output score for neuron, $k$, obtained by the linear combination of its inputs and weights.

During the training process, embeddings in the Embedding layer and weights in the remaining layers (LSTM and fully-connected) are learnt using backpropagation algorithm with the cross-entropy based loss functions \cite{goodfellowdlbook2016}.

\subsection {Feature extraction}

After the model has been trained, any integer encoded ME can be fed to it and the output after the global pooling layer, which is of size, 64, is obtained. This 64-dimensional output forms the feature vector for the given input integer encoded ME. 

Feature extraction using the deep learning model has been illustrated on the earlier mentioned example ME, in Table \ref{featext}.

\renewcommand{\arraystretch}{1.5}

\begin{table*}
\caption{Illustration of feature extraction from integer encoded MEs using deep learning architecture with the following model parameters: Embedding size, $e = 16$, Number of neurons in LSTM RNN, $t=64$.}
\begin{tabular}{|p{\textwidth}|}
\hline
Input LaTeX expression: \\
\textbf{\textbackslash sum\_\{i=a\}\^{ }b f(i)} ($\sum_{i=a}^b f(i)$). \\
Its integer encoding form \cite{ppkretrieval} (13 symbols) which is given as input to Embedding layer (Look-up table):\\
102, 1000, 1004, 201, 1004, 1001, 1002, 1004, 1003, 1004, 156, 1004, 157. \\ \hline
Embedding layer with embedding dimension, $e=16$,\\
{\bf Output:} 13 $\times$ 16 matrix \\ 
This matrix (13 symbols, as 13 time steps, each of dimension, 16) is fed as input to LSTM RNN. \\ \hline
LSTM RNN layer with $t=64$ neurons:\\
{\bf Output:} 13 $\times$ 64 matrix, 64 outputs (number of neurons in RNN layer) for each symbol.\\
This matrix, is fed as input to Global pooling layer. \\ \hline
{\bf Global pooling layer --} Applies pooling (minimum) operation on each column of the 13 $\times$ 64 matrix, to obtain a single vector of size, 64.\\ \hline\hline
{\bf Feature Extraction:} The 64-dimensional output from the global pooling layer serves as a feature vector for the given input integer encoded ME.\\ \hline
\end{tabular}
\label{featext}
\end{table*}
\subsection{Retrieval}
 As mentioned in the previous subsection, feature vectors are similarly computed for each database ME to generate a feature database. Feature vector of a query is then matched with the feature vectors of all the database MEs using the euclidean distance, and top $k$ matches are retrieved. 


As matching is based on the euclidean distance, it takes linear time, $O(t)$, where $t$ is the size of the feature vector. But LCS based matching has quadratic time complexity, $O(n^2)$ where $n$ is the number of symbols in the input integer encoded ME \cite{ppkretrieval}. If there are $T$ database MEs, the complete retrieval process using deep learning approach takes $O(Tt)$ complexity whereas LCS based retrieval method takes $O(Tn^2)$.

\section{Experimental results and discussion}
\label{results}
For the experimentation, we have applied our proposed deep learning model on our PACME LaTeX database \cite{pacme}. However, our model provides a general framework for the retrieval of any other unstructured text (like MathML, HTML etc.) by just defining the corresponding vocabulary. 

\begin{table*}[t]
\centering
\caption{Top 5 retrieved MEs for a query (shown in the first row) using deep learning and LCS based approaches.}
\label{res1}
\vspace{0.1cm}
\begin{tabular}{|c|c|c|}
\hline
S.No. & Deep learning approach & LCS based approach \\ \hline\hline
1 (Query) & \multicolumn{2}{c|}{$  \int (u+v) dx= \int udx+ \int vdx $}\\ \hline
2 & $  \int f(x) \pm g(x)dx= \int f(x)dx \pm  \int g(x)dx $ & $  \int cudx=c \int udx $
\\ \hline
3 & $  \int f(x) \pm h(x)dx= \int f(x)dx \pm  \int h(x)dx $
 & $  \int udv=uv- \int vdu $\\ \hline
4 &$  \int udv=uv- \int vdu $
 & $  \int xdy=xy- \int ydx $\\ \hline
5 &$  \int xdy=xy- \int ydx $
 & $  \int f(x) \pm g(x)dx= \int f(x)dx \pm  \int g(x)dx $\\ \hline
6 & $  \int  \text{csch}  \coth udu=- \text{csch} u+c $
& $  \int f(x) \pm h(x)dx= \int f(x)dx \pm  \int h(x)dx $\\ \hline
\end{tabular}
\end{table*}

In Table \ref{res1}, we have shown an ME query (Integral expression) in the first row. As discussed earlier, it is converted to the integer encoding format, fed to the trained deep learning model and then features are obtained (from the output of global pooling layer). These features are compared with those of database MEs using the euclidean distance and top five matches are retrieved. If the query is also present in the database, it is the best match (top one). In the second column of the table, we have shown top 5 retrieved MEs (excluding the query) using the above deep learning approach. For the sake of comparison, we have also shown the top 5 retrieved MEs for the same query using LCS based approach \cite{ppkretrieval} in the third column of the table. From this table, it can be observed that MEs retrieved using the deep learning approach are semantically similar. The retrieved MEs (especially the top two in the second and third rows) carry the following semantics: {\bf Integral of the sum of two terms is equal to the sum of integration of individual terms}. On the other hand, MEs retrieved using LCS based approach are syntactically similar based on the number of exact matches in a sequence. However, as mentioned earlier, LCS based matching has quadratic time complexity while deep learning approach takes only linear time.

\renewcommand{\arraystretch}{1.5}
\begin{table}
\centering
\caption{2D output heat maps of RNN layer for the query and its 5 retrieved MEs using deep learning approach, shown in Table \ref{res1}.}
\label{res2}
\vspace{0.1cm}
\begin{tabular}{|c|c|}
\hline
S.No. & Feature map \\ \hline\hline
1 (Query) & \includegraphics[scale=1]{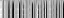}\\ \hline
2 & \includegraphics[scale=1]{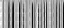}\\ \hline
3 & \includegraphics[scale=1]{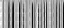}\\ \hline
4 & \includegraphics[scale=1]{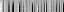}\\ \hline
5 & \includegraphics[scale=1]{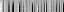}\\ \hline
6 & \includegraphics[scale=1]{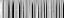}\\ \hline
\end{tabular}
\end{table}

Further, we have extracted the outputs of RNN layer (just before the global pooling layer) to inspect the features learnt by this layer. As the output of RNN layer is of size, ($l \times t$), these two-dimensional (2D) outputs are graphically represented in the form of heat maps \cite{goodfellowdlbook2016} which are shown in Table \ref{res2}. As already discussed, $l$ may be different for different MEs and $t$ is fixed. From the table, it can be observed that the heat maps of the retrieved MEs have similar patterns to that of the query ME. The successive global pooling layer  converts these 2D outputs (features of size, ($l \times t$)) into a one-dimensional output (features of size, ($1 \times t$)), which is ultimately used for the retrieval process. This experiment is helpful to visually analyze and explain the similarity of heat map patterns (explainability).

\begin{table*}
\centering
\caption{Top 5 retrieved MEs for a complex query (shown in the first row) using deep learning and LCS based approaches.}
\label{res3}
\vspace{0.1cm}
\begin{tabular}{|c|c|c|}
\hline
S.No. & Deep learning approach & LCS based approach \\ \hline\hline
1 (Query) & \multicolumn{2}{c|}{$  \int  \frac {dx}{x \sqrt {x^{2}+a^{2}}}= \frac {1}{a} \ln  \mid  \frac {x}{a+ \sqrt {a^{2}+x^{2}}} \mid  $} \\ \hline
2 &   $\int  \frac {dx}{x \sqrt {x^{2}-a^{2}}}= \frac {1}{a} \arccos  \frac {a}{ \mid x \mid },a>0 $ & $  \int  \frac {dx}{ \sqrt {a^{2}-x^{2}}}=- \frac {1}{a} \ln  \mid  \frac {a+ \sqrt {a^{2}-x^{2}}}{x} \mid  $ \\ \hline 
3 & $  \int  \frac {1}{u \sqrt {u^{2}-a^{2}}}du= \frac {1}{a} \sec ^{-1}( \frac {u}{a})+c $ & $  \int  \frac {dx}{ax^{2}+bx}= \frac {1}{a} \ln  \mid  \frac {x}{a+bx} \mid  $
\\ \hline
4 & $  \int  \frac {dx}{ \sqrt {a^{2}-x^{2}}}=- \frac {1}{a} \ln  \mid  \frac {a+ \sqrt {a^{2}-x^{2}}}{x} \mid  $ & $  \int  \frac {dx}{x \sqrt {x^{2}-a^{2}}}= \frac {1}{a} \arccos  \frac {a}{ \mid x \mid },a>0 $ \\ \hline
5 & $  \int  \frac {dx}{ \sqrt {a^{2}+x^{2}}}= \ln (x+ \sqrt {a^{2}+x^{2}}),a>0, $ & $  \int  \frac { \sqrt {a^{2}+x^{2}}}{x}dx= \sqrt {a^{2}+x^{2}}-a \ln  \mid  \frac {a+ \sqrt {a^{2}+x^{2}}}{x} \mid  $\\ \hline
6 & $  \int  \frac {dx}{ \sqrt {x^{2}-a^{2}}}= \ln  \mid x+ \sqrt {x^{2}-a^{2}} \mid ,a>0, $ & $  \int  \frac {dx}{a^{2}-x^{2}}= \frac {1}{2a} \ln  \mid  \frac {a+x}{a-x} \mid  $
\\ \hline
\end{tabular}
\end{table*}

In Table \ref{res3}, we have shown a complex ME query and its top 5 retrieved MEs (excluding the query) using both deep learning and LCS based approaches. From the second column of this table, it can be observed that MEs retrieved using deep learning approach are semantically similar. In each of these equations, left hand side (LHS) shows a function on which integration is applied while right hand side (RHS) shows the result of the integration. LHS has complex functions with fractions, quadratic expressions inside squareroot etc., while RHS has mathematical functions (like {\bf ln}, {\bf arccos}, {\bf sec${}^{-1}$} etc.), which are applied on fractional terms (especially, for the top three MEs in the second, third and fourth rows of the second column). For the sake of comparison, we have also shown the MEs retrieved using LCS based approach in the third column of the table. It can be observed that these MEs are syntactically similar based on the number of exact matches in a sequence.

From the above experiments, it can be concluded that deep learning based ME retrieval algorithm focuses on the semantic similarity. Further, this deep learning based retrieval framework is also flexible as any type of semantic notion can be incorporated by choosing a classification task based on the user requirements (like similar complexity, similar field etc.).

\section{Conclusions}
\label{conc}
In our present work, we have focused on deep learning approaches for ME retrieval based on semantic similarity. This semantic based retrieval approach is also useful for math plagiarism detection. Our deep learning model considers variable-length MEs without padding. This model has been initially trained on an auxiliary classification task based on the user requirements like complexity, field etc. After the training process, for all the database MEs, features are computed from the output of global pooling layer of the model and stored. For a given query, its features are similarly computed using the trained model. Query features are compared with the features of the database MEs using the euclidean distance and top $k$ MEs are retrieved. As matching is based on the euclidean distance, it has linear time complexity. In future, our model will be incorporated into a distributed learning framework with distributed ME databases.

\bibliographystyle{apalike} 
\bibliography{MathEmbeddingArxiv}

\end{document}